\documentclass[a4paper]{article}

\usepackage{INTERSPEECH2022}
\usepackage{amsmath,amsfonts,graphicx}
\usepackage{amssymb,textcomp,mathtools}
\usepackage{bm,upgreek,algorithm,hyperref}
\usepackage{multirow,booktabs,hhline,array}
\usepackage{cite,url,makecell,setspace}
\usepackage{xcolor}

\renewcommand{\vec}[1]{\bm{\mathrm{#1}}}
\def\thline{\noalign{\hrule height 1.0pt}}

\title{On the Use of Deep Mask Estimation Module for Neural Source Separation Systems}
\name{Kai~Li$^{\dagger}$$^{\ddagger}$\sthanks{Work done during internship at Tencent AI Lab.}, Xiaolin Hu$^\dagger$, Yi~Luo$^\ddagger$}
\address{$^\dagger$Department of Computer Science and Technology, BNRist, Tsinghua University, China \\
         $^\ddagger$Tencent AI Lab, Shenzhen, China}
\email{lk21@mails.tsinghua.edu.cn, xlhu@tsinghua.edu.cn, oulyluo@tencent.com}

\begin{document}
\maketitle
\begin{abstract}
Most of the recent neural source separation systems rely on a masking-based pipeline where a set of multiplicative masks are estimated from and applied to a signal representation of the input mixture. The estimation of such masks, in almost all network architectures, is done by a single layer followed by an optional nonlinear activation function. However, recent literatures have investigated the use of a deep mask estimation module and observed performance improvement compared to a shallow mask estimation module. In this paper, we analyze the role of such deeper mask estimation module by connecting it to a recently proposed unsupervised source separation method, and empirically show that the deep mask estimation module is an efficient approximation of the so-called \textit{overseparation-grouping} paradigm with the conventional shallow mask estimation layers.
\end{abstract}
\section{Introduction}
\label{sec:intro}
Among the existing neural network frameworks for source separation, the most widely-used configuration is the \textit{masking-based} separation pipeline where a set of multiplicative masks are estimated and applied to a representation of the mixture waveform. For Fourier-transform-based signal representations, the masks are defined in the time-frequency (T-F) domain, and multiple oracle or ideal masks can be defined as training objectives for a given system \cite{wang2014training, hummersone2014ideal, benaroya2003non, erdogan2015phase, williamson2016complex}. For learnable signal representations, the masks can be jointly optimized with the signal representations towards a single training objective \cite{venkataramani2018end, luo2019conv}.

The estimation of the multiplicative masks in most existing pipelines is done by a single output layer followed by an optional nonlinear activation function. For a mixture signal with $C$ target sources, a feature generated by a sequence of neural network modules is typically sent to $C$ fully-connected (FC) layers to generate the $C$ multiplicative masks \cite{huang2014deep, yu2017permutation, luo2020dual, subakan2021attention, zeghidour2021wavesplit, lam2021sandglasset,afrcnn}. Although such configuration achieves satisfying separation performance with many neural network architectures, recent works have explored the use of a deeper mask estimation module and observed consistent performance improvement \cite{luo2021rethinking}. The deeper mask estimation module, which was referred to as the \textit{SIMO-SISO} configuration in \cite{luo2021rethinking}, applied a single-input-multi-output (SIMO) module to generate $C$ features and a deep single-input-single-output (SISO) module on each of the feature to generate the multiplicative mask. However, why such deeper mask estimation module led to a better separation performance remains unclear.

In this paper, we analyze the role of such deeper mask estimation module in the neural source separation pipelines by connecting it to a recently proposed unsupervised source separation method, the mixture-of-mixtures method (\textit{MixIt}) \cite{wisdom2020unsupervised}. MixIt performs unsupervised source separation by mixing $K$ mixture signals to form a mixture-of-mixtures (MoMs) which contains $C\geq K$ target sources, and estimates $P\geq C$ outputs from the MoMs as the system outputs. The $P$ outputs are then \textit{grouped} or \textit{summed} into $K$ mixtures that best reconstructs the $K$ input mixtures used to create the MoMs. Since the number of outputs $P$ is always greater or equal to both the number of mixtures $K$ and the number of target sources $C$, such separation configuration can be viewed as an \textit{overseparation} paradigm. We show that when applied to the supervised training framework, the deeper SISO mask estimation module can be viewed as a simple replacement to the overseparation-grouping pipeline with the conventional single-layer mask estimation module. We also empirically find that (1) the performance of the overseparation-grouping pipeline can be improved by increasing the number of outputs $P$, and (2) a simple multilayer perceptron (MLP) mask estimation module can easily achieve on par performance as the overseparation-grouping pipeline in a more efficient way.

The rest of the paper is organized as follows. Section~\ref{sec:DMEM} introduces the overall design of the deep mask estimation module and shows how it is connected to the overseparation-grouping pipeline. Section~\ref{sec:config} describes the experiment configurations. Section~\ref{sec:discuss} presents the experiment results and discussions. Section~\ref{sec:conclusion} concludes the paper.

\begin{figure*}[!ht]
\centering
\includegraphics[width=0.8\textwidth]{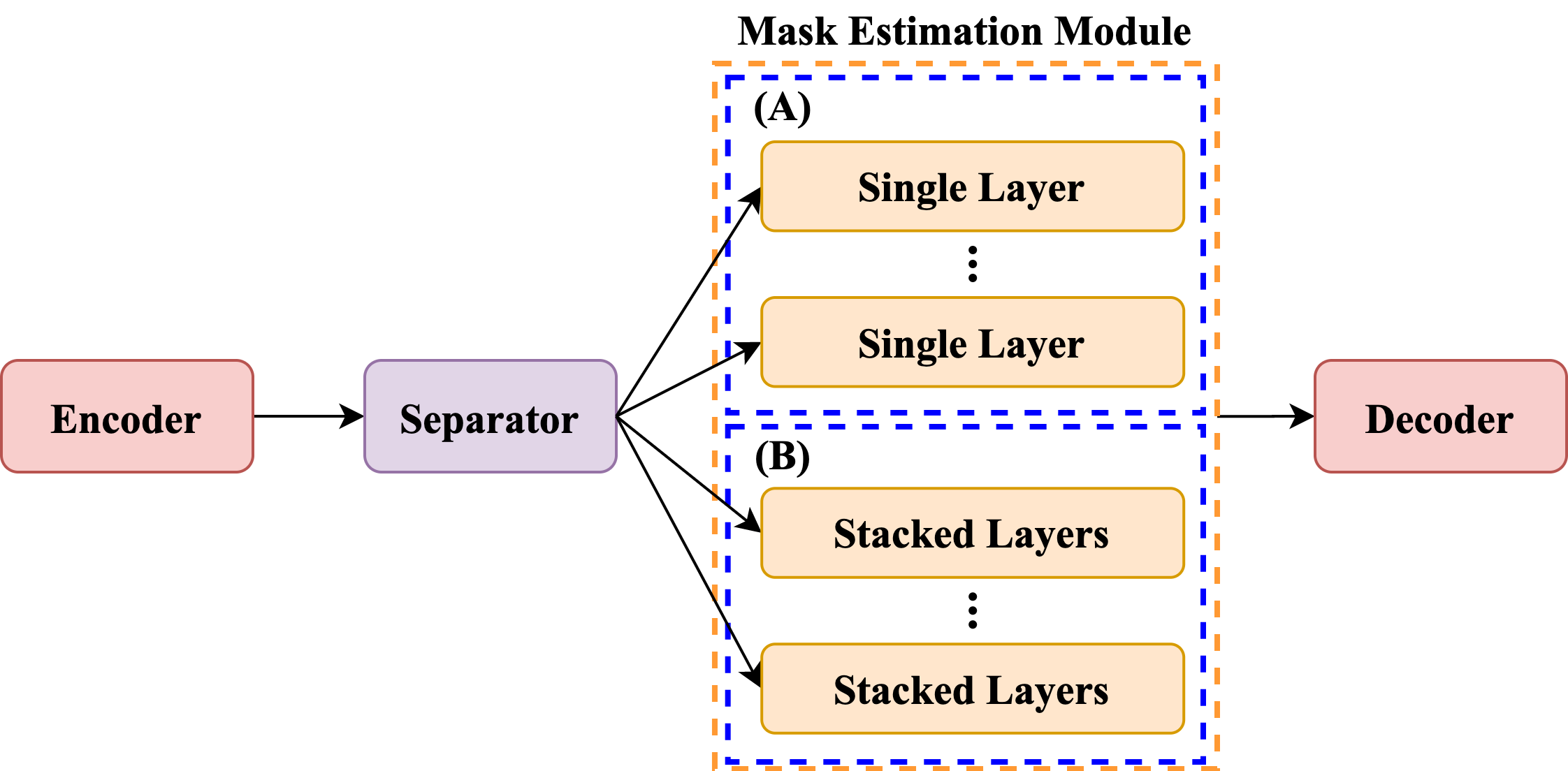}
\caption{Flowchart of a standard masking-based neural source separation pipeline. (A) A single layer for a shallow mask estimation module. (B) Stacked layers for a deep mask estimation module.}
\label{fig:overview}
\end{figure*}

\section{Deep Mask Estimation Module}
\label{sec:DMEM}
\subsection{System overview}

Figure~\ref{fig:overview} presents a standard masking-based neural source separation pipeline with different mask estimation modules. An encoder first encodes the mixture waveform to a latent representation, and the representation is sent to a separator to estimate $C$ multiplicative masks corresponding to the $C$ target sources, and a decoder reconstructs the waveforms from the $C$ masked mixture representations. Most existing pipelines use a shallow mask estimation module which typically consists of a single fully-connected (FC) layer with an optional nonlinear activation function (e.g. ReLU or Sigmoid), which is shown in Figure~\ref{fig:overview} (A).

Figure~\ref{fig:overview} (B) shows the simple modification to the conventional pipeline where multiple stacked layers are used in the mask estimation module. This modification contains the SIMO-SISO configuration in \cite{luo2021rethinking} which used multiple stacked dual-path RNN (DPRNN) layers \cite{luo2020dual} in the mask estimation module, while other types of network architectures can also be used for the stacked layers. Here we simply use a multi-layer perceptron (MLP) with a total of 3 layers with hyperbolic tangent (Tanh) as the nonlinear activation function for the first and second layers. The nonlinear activation function for the last layer is kept the same as the FC layer in the conventional pipeline.

\subsection{Connection to the Overseparation-grouping Pipeline}

The so-called \textit{overseparation-grouping} pipeline is derived from the \textit{MixIt} method, a recently proposed method for training unsupervised source separation networks. As described in Section~\ref{sec:intro}, MixIt creates an MoMs from $K$ mixture signals and estimates $P$ outputs via a neural network, where $P$ is set to be greater or equal than the total number of target sources $C$ (with $C\geq K$). The term \textit{overseparation} comes from the configuration of $P\geq C$. The $P$ outputs are then assigned to $K$ groups, and all outputs in the same group are summed to form an output mixture. The output assignment that best reconstructs all the $K$ input mixtures are used for backpropagation during training.

Although MixIt was proposed for unsupervised learning, we can easily adopt the overseparation-grouping paradigm in the supervised learning framework. Assume that each of the $K$ input mixture signals are single-source signals and $C=K$, the conventional mask estimation module thus contains $P\geq C$ FC layers in the mask estimation module. Denote the latent representation of the mixture as $\vec{S} \in \mathbb{R}^{N\times T}$ where $N$ is the feature dimension of the representation and $T$ is the number of frames, the input feature to the $P$ FC layers in the mask estimation module as $\vec{H} \in \mathbb{R}^{H\times T}$ where $H$ denotes the feature dimension of the separator module, the multiplicative masks as $\vec{M}_p \in \mathbb{R}^{N\times T}, p=1,\ldots, P$, then the calculation of the masks can be written as:
\begin{align}
    \vec{M}_p = f(\vec{W}_p\vec{H} + \vec{b}_p)
\end{align}
where $\vec{W}_p \in \mathbb{R}^{N\times H}$ and $\vec{b}_p \in \mathbb{R}^{N\times 1}$ are the weight and bias of the $p$-th FC layer, respectively, and $f(\cdot)$ represents the nonlinear activation function. Given that the masks are applied to $\vec{S}$ via Hadamard product, the grouping stage in MixIt is equivalent to the generation of a new multiplicative mask by summing over a set of masks:
\begin{align}
\label{eqn:grouping}
    \hat{\vec{M}}_k = \sum_{p \in \Pi_k} \vec{M}_p = \sum_{p \in \Pi_k} f(\vec{W}_p\vec{H} + \vec{b}_p)
\end{align}
where $\Pi_k$ denotes the indices in $k$-th group. Equation~\ref{eqn:grouping} represents a complicated nonlinear mapping which is defined by a sum of multiple nonlinear mappings, and we can approximate this mapping via another neural network:
\begin{align}
    \sum_{p \in \Pi_k} f(\vec{W}_p\vec{H} + \vec{b}_p) \approx g_k(\vec{H})
\end{align}
where $g_k(\cdot)$ is a nonlinear mapping for $k$-th target source defined by a neural network. In our modified pipeline, $g_k(\cdot)$ is defined by the 3-layer MLP in the deep mask estimation module. Note that each of the $C$ target sources has its own MLP for mask estimation. 

It is easy to see that when $f(x)=x$, i.e. no nonlinear activation function is used for mask estimation, such approximation is no longer necessary as $\sum_{p \in \Pi_k} f(\vec{W}_p\vec{H} + \vec{b}_p) \coloneqq \hat{\vec{W}}_p\vec{H} + \hat{\vec{b}}_p$ where $\sum_{p \in \Pi_k} \vec{W}_p \coloneqq \hat{\vec{W}}_p$ and $\sum_{p \in \Pi_k} \vec{b}_p \coloneqq \hat{\vec{b}}_p$. In this case the performance of the overseparation-grouping pipeline with any $P>C$ should be on par with that with $P=C$. This further shows that the approximation made by the deep mask estimation module is only valid when a nonlinear activation function is used for mask estimation. For systems where masks with unbounded entries are estimated and no nonlinear activation function is required \cite{hu2020dccrn, gu2021complex}, one can use alternative nonlinear activation functions such as the parametric ReLU (PReLU) \cite{he2015delving} or gated linear units (GLU) \cite{dauphin2017language} to make the deep mask estimation module effective.

\section{Experiment Configurations}
\label{sec:config}
\subsection{Dataset and Model Configurations}
\label{sec:datasets}

We use the widely-used WSJ0-2Mix, WSJ0-3Mix \cite{hershey2016deep} and WHAMR! \cite{maciejewski2020whamr} dataset for single-channel speech separation to validate the relationship between the modified deep mask estimation module and the overseparation-grouping pipeline. WSJ0-2mix and WSJ0-3mix dataset contain 30 hours of 4-second-long 8 kHz training data generated from the Wall Street Journal (WSJ0) si\_tr\_s set and 10 hours and 5 hours of validation and test data generated by using the si\_dt\_05 and si\_et\_05 sets, respectively. Each mixture is generated by randomly selecting speakers from the corresponding set and mixing them at a random relative signal-to-noise ratio (SNR) between -5 and 5 dB. WHAMR! extends the anechoic and noise-free WSJ0-2mix data by real-world noise and artificial reverberations.

\subsection{Model Configurations}
\label{sec:model}

We follow the standard configuration of DPRNN \cite{luo2020dual} in all models and use a 2~ms window size as the kernel size of the encoder and decoder. The number of feature dimension $N$ is set to 64. We set the total number of DPRNN blocks to 6 in all experiments, and we set the number of hidden units in all LSTM layers to 128. For the overseparation-grouping pipeline, the number of system outputs $P$ is set to 4, 8 and 16 for comparison, and the number of target sources $C$ is always set to 2. ReLU activation is used as the nonlinear activation function in the FC mask estimation layers. For the grouping stage, we use a \textit{deterministic} grouping strategy where the first $P/2$ outputs are summed to create the first target source, and the next $P/2$ outputs are summed to create the second target source. Note that this is different from the dynamic grouping strategy in MixIt where all possible combinations of the system outputs are calculated and compared to the $K$ input mixtures, and the reason we select this simplified grouping strategy is not only because this simplifies the training pipeline, but also because we empirically find that the two strategies lead to similar performance in our supervised separation experiments. For the MLP used in the modified pipeline, we adopt two MLPs with 16 and 64 hidden units, respectively, to compare the effect of different modeling capacity in $g(\cdot)$. ReLU activation is used for the last layer as the nonlinear activation function.

\subsection{Training and Evaluation}
\label{sec:param}

For training, we use negative SNR between the model output and the reverberant clean targets as the training objective. We use the Adam optimizer \cite{kingma2014adam} with the initial learning rate of 0.001, and we decay the learning rate by a factor of 0.5 if no best training model is found in three consecutive epochs. Gradient clipping by a maximum gradient norm of 5 is applied. The batch size is set to 4 for all experiments. We train the models until no best validation model is found in 15 consecutive epochs.

For evaluation, the scale-invariant signal-to-distortion ratio improvement (SI-SDRi) \cite{le2019sdr} and signal-to-distortion ratio improvement (SDRi) \cite{vincent2006performance} are selected to measure the speech separation performance. The model complexity is measured by both the number of parameters and the number of multiply-accumulate (MAC) operations \cite{whitehead2011precision} \footnote{https://github.com/Lyken17/pytorch-OpCounter}.

\section{Results and Discussions}
\label{sec:discuss}
\begin{table}[!htbp]
\centering
\small
\begin{tabular}{ccccc}
\thline
$P$ & $f(\cdot)$ & SI-SDRi (dB) & Param. (M) & MACs (G) \\
\hline
\multirow{2}{*}{2} & ReLU & 16.3 & \multirow{2}{*}{2.6} & \multirow{2}{*}{21.5} \\
 & -- & 15.9 & \\
 \hline
 \multirow{2}{*}{4} & ReLU & 17.4 & \multirow{2}{*}{2.6} & \multirow{2}{*}{21.6} \\
 & -- & 16.3 & \\
 \hline
 \multirow{2}{*}{8} & ReLU & 17.4 & \multirow{2}{*}{2.6} & \multirow{2}{*}{21.8} \\
 & -- & 16.4 & \\
 \hline
 \multirow{2}{*}{16} & ReLU & \textbf{17.8} & \multirow{2}{*}{2.7} & \multirow{2}{*}{22.2} \\
 & -- & 16.5 & \\
\thline
\end{tabular}
\caption{Separation performance of various overseparation-grouping configurations on the WSJ0-2mix dataset. MACs are calculated on a 4-second-long input.}
\label{tab:oversep}
\end{table}

\begin{table*}[!htp]
\centering
\begin{tabular}{ccccccc}
\toprule
\multirow{2}{*}{Model} & \multicolumn{2}{c}{WSJ0-2Mix} & \multicolumn{2}{c}{WHAMR!} & \multirow{2}{*}{Params (M)} & \multirow{2}{*}{MACs (G)} \\
& SI-SDRi (dB) & SDRi (dB) & SI-SDRi (dB) & SDRi (dB) & & \\
\midrule
Baseline (6 DPRNN blocks) & 16.3 & 16.5 & 10.0 & 11.0 & 2.6 & 21.5 \\
\qquad+$P=16$ & 17.8 & 18.0 & \textbf{10.7} & \textbf{11.6} & 2.7 & 22.2 \\
\qquad+MLP (S) & 17.5 & 17.7 & 10.3 & 11.3 & 2.6 & 21.5 \\
\qquad+MLP (L) & \textbf{18.0} & \textbf{18.1} & 10.6 & 11.5 & 2.6 & 21.5 \\
\hline
Baseline (9 DPRNN blocks) & 17.2 & 17.4 & 10.2 & 10.8 & 3.9 & 32.1 \\
Baseline (12 DPRNN blocks) & 17.3 & 17.6 & 10.4 & 11.0 & 5.2 & 42.8 \\
\bottomrule
\end{tabular}
\caption{Performance comparison of models with either overseparation-grouping or MLP-based deep mask estimation modules on the WSJ0-2mix dataset.}
\label{tab:deep}
\end{table*}

\subsection{Performance of Overseparation-grouping Pipeline}

We first present the separation performance of the overseparation-grouping pipeline. Table~\ref{tab:oversep} shows the performance of various choices of the number of mask estimation layers $P$ as well as the nonlinear activation function $f(\cdot)$. $P=2$ represents the baseline where no grouping is required. We first observe that for the models without a nonlinear activation function $f(\cdot)$ for mask estimation, increasing the number of outputs $P$ does not lead to significant improvement on separation performance. This matches our discussion in Section~2.2. On the other hand, we can see that setting $f(\cdot)$ to ReLU enables the model to significantly improve the separation performance as $P$ becomes larger. The results show that the overseparation-grouping pipeline is an effective method for improving the separation performance in supervised-training configuration.

\subsection{Performance of Deep Mask Estimation Module}

We then show that the deep mask estimation module is a replacement to the overseparation-grouping pipeline. Table~\ref{tab:deep} provides the performance comparison between the baseline model, baseline model with $P=16$ outputs, and baseline model with small (S) or large (L) MLP-based deep mask estimation modules. We find that the large MLP model, which contains 64 hidden units, achieves on par performance as the overseparation-grouping pipeline with 16 outputs in both WSJ0-2mix and WHAMR! dataset. Moreover, as the number of model parameters in the mask estimation layer in the overseparation-grouping pipeline is directly related to $P$, a large $P$ leads to not only a larger model size but also a higher model complexity. The deep mask estimation module is thus a simpler way to mimic the behavior of the overseparation-grouping paradigm at a lower computational cost, as a 3-layer MLP can be enough to achieve the same performance as a 16-output overseparation-grouping model. We further notice that the small MLP model with 16 hidden units has similar performance than the 8-output overseparation-grouping model, further showing that the model capacity of the mask estimation module can be directly compared to the number of outputs $P$ or the complexity of the mapping function defined in equation~\ref{eqn:grouping}. These results empirically prove that the deep mask estimation module can be viewed as an alternative or replacement of the overseparation-grouping pipeline to achieve a better separation performance than the conventional single-layer mask estimation module.

Given that performance of a system can also be improved by increasing the model size and capacity, we also conduct experiments on comparing the performance of a deeper separator and a deeper mask estimation module. The last two rows in Table~\ref{tab:deep} show the performance of the systems with 9 and 12 DPRNN blocks in the separator. We can observe that increasing the depth of the separator is able to improve the separation performance, while the performance of the 12-block system is only slightly better than the 9-block system. Moreover, both systems show worse performance by a simple deep mask estimation module even at a cost of a significantly larger amount of model parameters and MACs. This shows that when one is considering increasing the model size, it is always good to consider the proper place or module for the increase.

\begin{table}[!htbp]
\centering
\small
\begin{tabular}{ccc}
\thline
Model & SI-SDRi (dB) & SDIi (dB) \\
\hline
Baseline & 14.6 & 14.9 \\
\qquad+MLP (L) & \textbf{15.5} & \textbf{15.7} \\
\thline
\end{tabular}
\caption{Separation performance of shallow and deep mask estimation modules on the WSJ0-3mix dataset.}
\label{tab:3mix}
\end{table}

We further evaluate the effect of the deep mask estimation module on the three-speaker separation task. Table~\ref{tab:3mix} shows the performance of the baseline system and the deep mask estimation-based system. We can see that similar to the observation on the WSJ0-2mix dataset, where the deep mask estimation module achieves significantly better performance than the baseline with negligible additional computational cost. This shows that the deep mask estimation module is potentially helpful in various dataset configurations.

\section{Conclusion}
\label{sec:conclusion}
In this paper, we focused on the analysis of the role of a deep mask estimation module in masking-based neural source separation systems. Although most recent neural source separation systems contains a single-layer mask estimation module, there exists systems where a deep, multi-layer mask estimation module are applied to obtain a separation performance improvement. We provided an explanation to this phenomenon by connecting it to the so-called \textit{overseparation-grouping} pipeline, a pipeline extracted from a recently proposed unsupervised source separation method. We showed that while the overseparation-grouping pipeline was able to improve the separation performance, the deep mask estimation module can be viewed as a simple replacement of the overseparation-grouping pipeline at a lower computational cost. Moreover, we showed that using a deeper mask estimation module obtains more significant performance improvement than using a deeper separation module, indicating that properly increasing the model size can be important for the overall performance.

\bibliographystyle{IEEEbib}
\bibliography{ref}

\end{document}